\documentclass[11pt]{amsart}
\usepackage{hyperref}
\numberwithin{equation}{section} 
\def \dd{{\rm d}}
\def \DD{{\rm D}}
\begin{document}

\title[]{Gravitational singularities via acceleration:
The case of the Schwarzschild solution and Bach's gamma metric}%

\author{Salvatore Antoci}%
\address{Dipartimento di Fisica ``A. Volta'' and INFM, Pavia, Italy}%
\email{Antoci@fisicavolta.unipv.it}%
\author{Dierck-Ekkehard  Liebscher}%
\address{Astrophysikalisches Institut Potsdam, Potsdam, Germany}%
\email{deliebscher@aip.de}%
\author{Luigi Mihich}%
\address{Dipartimento di Fisica ``A. Volta'' and INFM, Pavia, Italy}%
\email{Mihich@fisicavolta.unipv.it}%
\keywords{General relativity - gamma metric - Schwarzschild
solution}%
\begin{abstract}
The so called gamma metric corresponds to a two-parameter family
of axially symmetric, static solutions of Einstein's equations
found by Bach. It contains the Schwarzschild solution for a
particular value of one of the parameters, that rules a
deviation from spherical symmetry. \\
It is shown that there is invariantly definable singular behaviour
beyond the one displayed by the Kretschmann scalar
when a unique, hypersurface orthogonal, timelike Killing vector exists.
In this case, a particle can be defined to be at rest when its world-line
is a corresponding Killing orbit. The norm of the acceleration on such an orbit
proves to be singular not only for metrics that deviate from Schwarzschild's metric,
but also on approaching the
horizon of Schwarzschild metric itself, in
contrast to the discontinuous behaviour of the curvature scalar.
\end{abstract}
\maketitle
\section{Introduction}
In the early days of general relativity there was little doubt
that in the new theory the Christoffel symbols had the role of
``components'' of the gravitational field
\cite{Einstein15b,Einstein16}. Einstein was very explicit on
this point: at page 802 of his paper \cite{Einstein16} on the
foundations of general relativity, he says:
\begin{quotation}
Verschwinden die $\Gamma^\tau_{\mu\nu}$, so bewegt sich der Punkt
geradlinig und gleichf\"ormig; diese Gr\"o\ss en bedingen also die
Abweichung der Bewegung von der Gleich\-f\"ormigkeit. Sie sind die
Komponenten des Gravitationsfeldes.\footnote[1]{If the
$\Gamma^\tau_{\mu\nu}$ vanish, the point moves uniformly on a
straight line; therefore these quantities cause the deviation of
the motion from uniformity. They are the components of the
gravitational field. (It should be added that from the projective
point of view a straight line is defined by being element of any
set of lines that intersect at most once and that are determined
by two distinct events. Hence any set of coordinate lines will do.
The ambiguity of this set is usually reduced by choosing a set of
geodesic lines, at the expense of the restriction of the
intersection axiom to only local validity.)}
\end{quotation}
This identification can be deemed satisfactory if only gravitation and
inertia are considered: it accounts for the shifty role that the
gravito-inertial field plays in the geodesic equations of motion
for a pole test particle whose four-velocity is $u^i$:
\begin{equation}\label{1.1}
\frac{\dd u^i}{\dd s}+\Gamma^i_{kl}u^ku^l=0.
\end{equation}

The coefficients of the affine connection are not tensorial. Hence
any identification with gravitation or inertia depends on the
choice of the coordinate system. For particular world lines
defined by the solution itself, this draw-back may be partly
reduced, as we shall see later. Equation (\ref{1.1}) can be
however read as a particular case of the law
\begin{equation}\label{1.2}
\frac{\dd u^i}{\dd s}+\Gamma^i_{kl}u^ku^l=a^i.
\end{equation}
This is the general relativistic extension of Newton's second law;
it can be derived from the conservation identities of the theory
\footnote[2] {For the case of an electrically charged test particle see
{\em e.g.} \cite{PU55}.},
and claims that the force per unit mass $a^i$ exerted by non-gravitational
fields on the pole test particle is balanced by the
gravito-inertial pull (per unit mass) expressed by the left-hand
side of (\ref{1.1}). Therefore, in order to provide a definition of
the gravito-inertial force, it is not enough to know the metric
$g_{ik}$, hence the Christoffel symbols: as stressed by Whittaker
\cite{Whittaker35}, ``in general relativity the gravitational
force, as measured by any observer, depends not only on the
observer's position but also on his velocity and acceleration'',
being in fact given by the four-vector $a^i$ of equation
(\ref{1.2}). The relativist of the present day, convinced of
the exclusive role that invariant entities play in general relativity,
may feel some relief in noticing that the gravitational pull,
defined by Einstein in terms of noninvariant quantities, according to
Whittaker is accounted for by a tensorial entity,
associated with the world-line followed by the pole test particle.
Furthermore, in the cases when this world-line for some reason,
like some symmetry of the metric, has not a contingent character, but
can be uniquely chosen through an invariant definition, the norm
\begin{equation}\label{1.3}
\alpha\equiv(-a_ia^i)^{1/2}
\end{equation}
of the four-acceleration will provide a fully invariant
and maybe physically relevant measure of the otherwise
elusive gravito-inertial force.\par
However, the evident merits of the definition of the
gravito-inertial pull adopted by Whittaker were not considered by Synge
\cite{Synge37} to be sufficient for overcoming the alleged
drawback, that the gravito-inertial force defined in this way happens
to vanish when the pole test particle undergoes geodesic motion.
Moreover, a change in attitude appeared to him necessary: in
Einstein's theory only relative kinematic measurements
involving nearby particles are permitted in general, hence one must
renounce the unattainable goal of determining absolutely the force
acted by the gravitational field on any particle, and must be content
with a differential law that only allows for the comparison of the
gravitational pull acting at adjacent events. Let us forget about
nongravitational forces and their balance with the gravito-inertial ones:
we consider two pole test particles executing geodesic motion, and
imagine that their world-lines $L$, $M$ be very close to each other.
If $\eta^i$ is the infinitesimal displacement vector drawn perpendicular
to $L$ from a point $A$ on $L$ to a point $B$ on $M$, the acceleration
of $B$  relative to $A$ is defined as the vector
\begin{equation}\label{1.4}
f^i=\frac{\DD^2\eta^i}{\dd s^2},
\end{equation}
where $\DD/\dd s$ indicates absolute differentiation and $\dd s$
is the infinitesimal arc length of the geodesic $L$ measured at $A$.
But Synge himself \cite{Synge34},\cite{Synge60} has proved that
\begin{equation}\label{1.5}
\frac{\DD^2\eta^i}{\dd s^2}+R^i_{~jkl}u^j\eta^ku^l=0,
\end{equation}
where $R^i_{~jkl}$ is the Riemann tensor and $u^i$ is the
four-velocity of the particle at $A$. By appealing to Newton
one then postulates that the excess of the gravitational force
at the event $B$ over the gravitational force at the event $A$ is
naturally defined (for unit test masses) to be the acceleration
of $B$ relative to $A$. Therefore one finds \cite{Synge37} that
the excess of the gravitational pull is given by
\begin{equation}\label{1.6}
f^i=-R^i_{~jkl}u^j\eta^ku^l.
\end{equation}
Whittaker's definition of the gravitational pull through the
four-acceleration of a single pole test particle in arbitrary motion
is superseded by a relative definition, in terms of the relative
four-acceleration of two adjacent pole test particles both executing
a geodesic motion. Whittaker's definition looks like the prolongation,
within general relativity, of a line of thought that can be traced back
to Newton and d'Alembert, and gives the acceleration four-vector a
fundamental physical role. Synge's definition of the relative pull is
instead one of the first attempts at building the
physical interpretation of the geometric structure of general
relativity by availing as few as possible of concepts inherited
from the physics of the past, and happens to give a central
role to the Riemann tensor itself. One should not think of this change
of attitude as an abrupt occurrence in the development of general
relativity: despite an eloquent counterexample
\cite{Levi-Civita18},\cite{Schucking85} exhibited by
Levi-Civita, the idea that a ``true''
gravitational field must necessarily entail a
nonvanishing Riemann tensor is already present in Eddington's
book of 1924. In that book Eddington \cite{Eddington24} proposes to tell
apart the fake gravitational waves, {\em i.e.} mere undulations of the
coordinate system that can propagate with the ``speed of thought'',
from the alleged true ones by looking at the behaviour of the Kretschmann
scalar $R^i_{~jkl}R_i^{~jkl}$. Whittaker's definition of the
gravitational pull is however still alive and well
in a paper written
by Rindler \cite{Rindler60} in 1960. Nevertheless, a quiet revolution has
indeed occurred in general relativity if, in the section entitled
``Gravitational field'' of his essay \cite{Synge66} in honour of the
geometer Hlavat\'y, Synge could eventually assert that the
wise plan is to forget about Newton's arrow and say
``gravitational field = curvature of space-time''.\par One cannot help
noticing that this assertion, of course impressive and general in
character, is much vaguer than the precise and somewhat technically
unwieldy definition (\ref{1.6}) of the relative gravitational
pull given by Synge in 1937. Nevertheless this assertion,
despite Levi-Civita's counterexample \cite{Levi-Civita18}, has
entered the mind habit of the relativist: whenever asked
about the properties of the gravitational field associated with
a given metric, in particular about its possible singularities,
one immediately thinks to the Riemann tensor or, better, to
the invariants that can be constructed from it.
\section{The case of the gamma metric}
Of course, the identification of a geometric entity of
Einstein's general relativity with a physical one has not to
be decided {\em a priori} on the basis of some
preconception about the best way for building theoretical
physics; a study of how the proposed identification works
in clear-cut examples provided by the
theory is a necessary requisite for settling such issues.
In the present paper we shall compare the insight in the behaviour
of the singularities of the gravitational field that can be obtained
through the Whittaker definition and through the Riemann tensor
approach in the particular case of the so called gamma metric.\par
The latter is one of the axially symmetric, static
solutions \cite{BW22} calculated in 1922 by Bach, who availed of
the general method \cite{Weyl17},\cite{Levi-Civita19}
found by Weyl and by Levi-Civita.
Despite the nonlinear structure of Einstein's equations, Weyl
succeeded in reducing the axially symmetric, static problem to
quadratures through the introduction of his ``canonical
cylindrical coordinates''. Let $x^0=t$ be the time coordinate,
while $x^1=z$, $x^2=r$ are the coordinates in a meridian half-plane,
and $x^3=\varphi$ is the azimuth of such a half-plane; the adoption
of Weyl's canonical coordinates allows writing the line
element of a static, axially symmetric field {\it in vacuo} as:
\begin{eqnarray}\label{2.1}
\dd s^2=e^{2\psi}dt^2-\dd\sigma^2,\;\\\nonumber
e^{2\psi}\dd\sigma^2
=r^2\dd\varphi^2+e^{2\gamma}(\dd r^2+\dd z^2);\nonumber
\end{eqnarray}
the two functions $\psi$ and $\gamma$ depend only on $z$ and $r$.
Remarkably enough, $\psi$ must fulfil the "Newtonian potential'' equation
\begin{equation}\label{2.2}
\Delta\psi=\frac{1}{r}\left\{\frac{\partial(r\psi_z)}
{\partial z}
+\frac{\partial(r\psi_r)}{\partial r}\right\}=0,
\end{equation}
where $\psi_z$, $\psi_r$ are the derivatives with respect to $z$ and to
$r$ respectively, while $\gamma$ is obtained by solving the system
\begin{equation}\label{2.3}
\gamma_z=2r\psi_z\psi_r,\;\gamma_r=r(\psi^2_r-\psi^2_z);
\end{equation}
due to the potential equation (\ref{2.2})
\begin{equation}\label{2.4}
\dd\gamma=2r\psi_z\psi_r\dd z+r(\psi^2_r-\psi^2_z)\dd r
\end{equation}
happens to be an exact differential.\par
The particular Bach's metric we are interested in is defined by
choosing for $\psi$ the potential that, in Weyl's ``Bildraum'',
is produced by a thin massive rod of constant linear density
$\sigma=k/2$ lying on the symmetry axis, say, between $z=z_2=-l$ and
$z=z_1=l$. One finds:
\begin{equation}\label{2.5}
\psi=\frac{k}{2}\ln\frac{r_1+r_2-2l}{r_1+r_2+2l},
\end{equation}
where
\begin{equation}\label{2.6}
r_i=[r^2+(z-z_i)^2]^{1/2},\;i=1,2.
\end{equation}
By integrating equations (\ref{2.3}) and by adjusting an integration
constant so that $\gamma$ vanish at the spatial infinity one obtains:
\begin{equation}\label{2.7}
\gamma=\frac{k^2}{2}\ln\frac{(r_1+r_2)^2-4l^2}{4r_1r_2}.
\end{equation}
The resulting metric is asymptotically flat at spatial infinity
and its components are everywhere regular, with the exception of
the segment of the symmetry axis for which $z_2\leq z\leq z_1$,
for any choice of the parameters $l$ and $k$, assumed here
to be positive.\par
It may be convenient \cite{Zipoy66} to express the line element
in spheroidal coordinates by performing, in the meridian
half-plane, the coordinate transformation \cite{CJ73}:
\begin{equation}\label{2.8}
\varrho=\frac{1}{2}(r_1+r_2+2l),\;\cos\vartheta=\frac{r_2-r_1}{2l}.
\end{equation}
Then the interval takes the form
\begin{eqnarray}\label{2.9}
ds^2=\left(1 - \frac{2l}{\varrho}\right)^k\dd t^2
-\left(1 - \frac{2l}{\varrho}\right)^{-k}\cdot\\\nonumber
\left[\left(\frac{\Delta}{\Sigma}\right)^{k^2-1}\dd\varrho^2
+\frac{\Delta^{k^2}}{\Sigma^{k^2-1}}\dd\vartheta^2
+\Delta\sin^2\vartheta\dd\varphi^2\right],
\end{eqnarray}
where
\begin{equation}\label{2.10}
\Delta=\varrho^2 - 2l\varrho,\;
\Sigma=\varrho^2-2l\varrho+l^2\sin^2\vartheta.
\end{equation}
We notice that when $k=1$ the metric reduces to Schwarz\-schild's
spherically symmetric one. It does so in the strict sense,
{\em i.e.} it is in one-to-one correspondence with
Schwarzschild's original solution \cite{Schwarzschild16},
not with the ``Schwarzschild metric'' of all the manuals and
research papers, that was actually found by Hilbert \cite{Hilbert17}.
The latter metric would be retrieved from (\ref{2.9})
with $k=1$ if the radial coordinate $\varrho$ were allowed the range
$0<\varrho<\infty$ while, due to (\ref{2.8}), the allowed
values of $\varrho$ are presently in the range $2l<\varrho<\infty$.\par
Let us now investigate the singularities of the gravitational
field of the gamma metric by adopting the point of view
according to which ``gravitational field = curvature of space-time''.
We shall explore the singular behaviour of the Riemann tensor in
an invariant way through the Kretschmann scalar $K=R^i_{~jkl}R_i^{~jkl}$.
The calculation and the study of this scalar has been done long
ago \cite{CJ73} by Cooperstock and Junevicus, and was recently
repeated by Virbhadra. We quote here his result \cite{Virbhadra96},
expressed with spheroidal coordinates:
\begin{equation}\label{2.11}
K=\frac{16l^2k^2N}
{\varrho^{2(k^2+k+1)}(\varrho-2l)^{2(k^2-k+1)}\Sigma^{3-2k^2} },
\end{equation}
with
\begin{eqnarray}\nonumber
&N=l^2 \sin^2\theta\cdot\\\nonumber&\left[3lk(k^2+1)(l-\varrho)
+k^2(4l^2-6l\varrho+3\varrho^2)+l^2(k^4+1)\right]\\\label{2.12}
&+3\varrho[(k+1)l-\varrho]^2(\varrho-2l).
\end{eqnarray}
We do not need a minute analysis of the function
$K(\rho,\vartheta,k)$ to decide that, for monitoring the
singularities of the gravitational field in the gamma metric, the
Kretschmann scalar is a discontinuous
parameter. It is sufficient to
examine the behaviour of $K$ in the neighbourhood of $k=1$, {\em
i.e.} for small axially symmetric deviations from spherical
symmetry. By studying the zeroes of both the numerator and the
denominator of (\ref{2.11}) one is then confronted with the
following situation. For all the values of $\vartheta$, the
denominator vanishes when $\rho\rightarrow 2l$, while the values
of $\varrho$ at which the zeroes of the numerator occur depend on
both $k$ and $\vartheta$. As a consequence in the neighbourhood of
$k=1$ the Kretschmann scalar always diverges \cite{CJ73} for
$\rho\rightarrow 2l$, provided that $k\neq 1$. When $k=1$, {\em
i.e.} when the metric reduces to Schwarzschild's one, both the
numerator and the denominator tend to zero as $\rho\rightarrow 2l$
for all $\vartheta$, and fatefully they do so in such a way that
the limit value of the Kretschmann scalar happens to be finite at
the ``Schwarzschild radius''.\par  It is evident that, since the
slightest deviation from spherical symmetry restores the
divergence of the Kretschmann scalar for $\rho\rightarrow 2l$,
the Kretschmann scalar is a discontinuous function on the set of Bach metrics
defined by equations (\ref{2.5}) and (\ref{2.7}). This behaviour
complies, in the particular case of the gamma metric, with a well known
theorem on the event horizons in static vacuum spacetimes \cite{Israel67}.
\par Let us adopt now the
viewpoint on the gravitational pull considered by Whittaker
\cite{Whittaker35}. Since the gamma metric is static, Whittaker's
approach allows in this case for a fully invariant and physically
transparent treatment. In fact, not only the norm (\ref{1.3}) of
the four-acceleration is a scalar, but also an invariantly
prescribed and physically privileged choice of the world lines of
the test particles is possible. This choice is not
arbitrarily imposed on the manifold, but is dictated by the manifold
itself in a purely local way. For choosing these world lines one
can in fact avail of the symmetries of the gamma
metric, and adopt the unique timelike Killing congruence that not only enjoys the
Killing property, but is also hypersurface orthogonal. Either in
canonical or in spheroidal coordinates this invariantly
distinguished congruence is identified by the constancy of the
spatial coordinates. In this case Whittaker's proposal both
fulfils the general requirement, that in general relativity the
relevant physical properties must be definable in invariant form,
and possesses a straightforward meaning, in keeping with
time-honoured ideas of pre-relativistic physics \cite{AL2001}. The
norm $\alpha$ of the four-acceleration measures in fact the
strength of the gravitational pull exerted on a pole test particle
of unit mass kept at rest in the given field.\par For the static,
axially symmetric interval of equation (\ref{2.1}) the
acceleration four-vector $a^i$ of a test particle kept at rest
($z$, $r$, $\varphi$ constant) has the nonvanishing components
\begin{equation}\label{2.13}
a^1=\exp{(2\psi-2\gamma)}\frac{\partial\psi}{\partial z},\;
a^2=\exp{(2\psi-2\gamma)}\frac{\partial\psi}{\partial r},
\end{equation}
hence its squared norm, in the particular case of the gamma
metric, reads
\begin{eqnarray}\label{2.14}
&\alpha^2
=\frac{16k^2l^2}{(4r_1r_2)^{1-k^2}(r_1+r_2-2l)^{1-k+k^2}
(r_1+r_2+2l)^{1+k+k^2}},
\end{eqnarray}
when referred to Weyl's canonical coordinates, and
\begin{eqnarray}\label{2.15}
&\alpha^2=\frac{16k^2l^2}{[4(\varrho-l)^2-4l^2\cos^2\vartheta]^{1-k^2}
(2\varrho-4l)^{1-k+k^2}
(2\varrho)^{1+k+k^2}},
\end{eqnarray}
when expressed as a function of the spheroidal coordinates
of equation (\ref{2.8}). For any value of $k<2$ and for all
$\vartheta$ the norm $\alpha$ happens to grow without limit as
$\varrho\rightarrow 2l$. At variance with what occurs with the
Kretschmann scalar, no discontinuous behaviour of $\alpha$ is noticed
when crossing the value $k=1$, for which the gamma metric acquires
spherical symmetry.
\section{Conclusion}
In the case of the gamma metric two definitions of the
gravitational field were considered with respect to their
use as indicators of physically meaningful singularities.
The behaviour of the Kretschmann scalar is
continuous only in carefully chosen
sets of solutions; in the set of gamma metrics, it is not.
On the contrary, the norm $\alpha$ of the four-acceleration
of a test particle kept at rest, that according to Whittaker should
provide the strength of the gravitational field measured on a unit mass
at rest in this static metric, appropriately vanishes for
$\varrho\rightarrow\infty$, is everywhere finite for
$2l<\varrho<\infty$, and uniformly diverges
in the limit $\varrho\rightarrow 2l$ for all the values of
the parameter $k$ in a suitable neighbourhood of $k=1$.\par
    In the existing literature the finite value of the Kretschmann
scalar at the ``Schwarzschild radius'' is adduced as an argument
for the viability of the program of analytic extension
\cite{Kruskal60},\cite{Szekeres60} of the Schwarzschild
solution.
The existence of this extension is an isolated property, and
it does not change the singular behaviour of the
invariantly defined Whittaker's force.\par
The uniformly divergent behaviour of the norm $\alpha$ when
$\varrho\rightarrow 2l$ confirms the wisdom of
Schwarzschild's deliberate choice \cite{Schwarzschild16}
to remove to the ``Nullpunkt'', {\em i.e.} to the border of the
manifold considered by him, the singular two-surface that
brings his name.\par

\end{document}